\documentclass[twocolumn]{aastex6}
\pdfoutput=1 
\usepackage{amsmath,amstext}
\usepackage[T1]{fontenc}
\usepackage{apjfonts} 
\usepackage[figure,figure*]{hypcap}
\usepackage{affils}
\usepackage{microtype}

\usepackage{xcolor}

\newcommand{\apx}{\ensuremath{\sim}}
\newcommand{\dd}{\ensuremath{\text{d}}}
\newcommand{\Lnil}{\ensuremath{L_{\nu,\text{iso,lim}}}}
\newcommand{\pergpcperyr}{Gpc$^{-3}$\,yr$^{-1}$}

\newcommand{\repeater}{FRB~121102}

\shorttitle{FRB and Magnetar Rates and Host Galaxy Demographics}
\shortauthors{Nicholl et al.}

\begin{document}

\title{Empirical constraints on the origin of fast radio bursts: volumetric rates and host galaxy demographics as a test of millisecond magnetar connection}

\DeclareAffil{cfa}{Harvard-Smithsonian Center for Astrophysics, 60 Garden Street, Cambridge, Massachusetts 02138, USA; \href{mailto:matt.nicholl@cfa.harvard.edu}{matt.nicholl@cfa.harvard.edu}}
\DeclareAffil{columbia}{Columbia Astrophysics Laboratory, Columbia University, New York, NY 10027, USA}

\affilauthorlist{M.~Nicholl\affils{cfa},
P.~K.~G.~Williams\affils{cfa},
E.~Berger\affils{cfa},
V.~A.~Villar\affils{cfa},
K.~D.~Alexander\affils{cfa},
T.~Eftekhari\affils{cfa},
B.~D.~Metzger\affils{columbia}
}

\begin{abstract}
The localization of the repeating fast radio burst (FRB)\,121102 to a low-metallicity dwarf galaxy at $z=0.193$, and its association with a luminous quiescent radio source, suggests the possibility that FRBs originate from magnetars, formed by the unusual supernovae that occur in such galaxies. We investigate this possibility via a comparison of magnetar birth rates, the FRB volumetric rate, and host galaxy demographics.  We calculate average volumetric rates of possible millisecond magnetar production channels such as superluminous supernovae (SLSNe), long and short gamma-ray bursts (GRBs), and general magnetar production via core-collapse supernovae.  For each channel we also explore the expected host galaxy demographics using their known properties. We determine for the first time the number density of FRB emitters (the product of their volumetric birthrate and lifetime), $R_{\rm FRB}\tau\approx 10^4$\,Gpc$^{-3}$, assuming that FRBs are predominantly emitted from repetitive sources similar to \repeater\ and adopting a beaming factor of 0.1.  By comparing rates we find that production via rare channels (SLSNe, GRBs) implies a typical FRB lifetime of \apx30--300~yr, in good agreement with other lines of argument. The total energy emitted over this time is consistent with the available energy stored in the magnetic field. On the other hand, any relation to magnetars produced via normal core-collapse supernovae leads to a very short lifetime of \apx0.5~yr, in conflict with both theory and observation. We demonstrate that due to the diverse host galaxy distributions of the different progenitor channels, many possible sources of FRB birth can be ruled out with $\lesssim 10$ host galaxy identifications. Conversely, targeted searches of galaxies that have previously hosted decades-old SLSNe and GRBs may be a fruitful strategy for discovering new FRBs and related quiescent radio sources, and determining the nature of their progenitors.
\end{abstract}


\section{Introduction}

Fast radio bursts (FRBs) are bright flares of coherent radio emission with millisecond durations and large dispersion measures ($\gtrsim 300$ pc cm$^{-3}$), well in excess of the expected Milky Way contribution. They were originally discovered in archival data by \citet{lbm+.2007} and are now regularly detected \citep{Keane+12, Thornton+13, Spitler+14, Ravi+15, Petroff+16, Champion+16}.  However, their origin remains poorly understood, in no small part because of a lack of precise localizations. The first claim of a subarcsecond spatial localization by \citet{kjb+.2016} was subsequently refuted by \citet{wb.2016}. The discovery of \repeater\ by \citet{sch+.2014} and its subsequent recognition as a repeating source with multiple outbursts \citep{ssh+.2016} were watershed moments in FRB science. \repeater\ was recently precisely localized by \citet{clw+.2017} to a host galaxy at a redshift of $z=0.1927$ \citep{tbc+.2017}. A VLBI localization and study by \citet{mph+.2017} revealed a luminous quiescent radio source coincident with the FRB, offset from the host optical emission centroid. The existence of a repeater rules out any cataclysmic channel for this FRB, and perhaps FRBs in general, but the luminosity of the bursts ($\nu L_\nu \sim 10^{38}$\,erg\,s$^{-1}$) still requires energetic events.

The host galaxy of \repeater\ was determined to be a dwarf galaxy with an absolute magnitude of $M_r\approx -17$ and a low metallicity of $12+{\rm log(O/H)}\lesssim 8.4$ \citep{tbc+.2017}. This would be surprising if FRBs simply traced normal stellar populations, in which case they would track stellar mass or star formation and hence occur in more luminous and massive galaxies. \citet{tbc+.2017} noted that similar dwarf galaxies also host hydrogen-poor superluminous supernovae (SLSNe) and long gamma-ray bursts (LGRBs). These transients are thought to be associated with the birth of neutron stars with dipole fields $B\sim 10^{14}$\,G and millisecond spin periods, termed millisecond magnetars \citep[e.g.][]{kas2010,met2015,nic2017b}, although in the case of LGRBs black hole engines are more typically assumed \citep{woo1993,macf1999}. Based on observations in the Galaxy, it appears that roughly $10\%$ of normal core-collapse supernovae (CCSNe) also produce magnetars, here termed `classical' magnetars, which may differ from the population inferred for SLSNe \citep{kou1998,gill2007}. It is not clear whether classical magnetars are born rapidly rotating, and their anomalous X-ray and gamma-ray activity seems to require a complicated magnetic field structure. In any case, less than 10\%\ of CCSNe occur in galaxies fainter than $M\approx -17$ \citep{sve2010}, so the host galaxy of \repeater\ makes a normal CCSN origin less compelling for this event.

Several works have attributed FRBs to young neutron stars or magnetars \citep{Kulkarni+15, Katz16, Cordes&Wasserman16, Lyutikov+16, Popov&Pshirkov16, Yang+16, Kumar+17}, possibly embedded within the ejecta shells of SN remnants \citep{Connor+16, Piro16, Murase+16,wax2017}. Following the localization of \repeater, \citet{mbm.2017} presented a magnetar model that consistently explains the properties of the FRBs,  the quiescent radio source, and the origin in a dwarf galaxy.  In this model the quiescent source is associated with a $\sim$\,decades-old supernova remnant from a SLSN or LGRB that created a millisecond magnetar, which then powers the repeating FRBs. The nebula could be energized by the magnetar or by interaction with the surrounding medium.  The key requirement setting the minimum age is that the ejecta have expanded sufficiently that they become transparent to the FRB emission, while the maximum age is set by the requirement not to overproduce the size of the quiescent radio source. Subsequent analyses have supported this broad picture \citep[e.g.][]{bel2017,lyu2017,piro2017}, although they invoke alternative mechanisms to power the quiescent radio source \citep{bel2017}. Flares from much older magnetars (millisecond or classical) are also a plausible model for the source of FRBs, as the magnetic energy can provide the necessary luminosity, and allow repetition \citep[e.g.][]{pop2013,lyu2014,pen2015,bel2017}. However, the quiescent radio source associated with \repeater\ may favor a younger magnetar than considered in most previous works.

If FRBs result from the magnetar remnants of rare explosions (SLSNe and/or LGRBs) or more common events (binary neutron star mergers, core-collapse SNe) this should be reflected in the event rate and host galaxy demographics.  In particular, the magnetar birth rate from each channel can be compared to the volume density of FRBs and the timescale over which they repeat.  Formation via rare channels will require a longer active lifetime compared to formation via common channels, and this can be compared to physical arguments about the FRB lifetimes (e.g., \citealt{mbm.2017}).  The host galaxy demographics will be similarly impacted in various formation scenarios.  

Here we carry out this analysis through an investigation of various FRB progenitor formation channels.  We furthermore study the expected demographics of FRB host galaxies in each scenario and the implications for future localizations in dwarf galaxies (or otherwise).  The outline of the paper is as follows. First we derive volumetric rates for events that can form magnetars (SLSNe, GRBs, and CCSNe; \S\ref{s:rates}). We then constrain the FRB repeater birth rate and lifetime, using \repeater\ for guidance, and compare these results to the magnetar birth rate and expected timescales (\S\ref{s:frbs}). We also use these results to constrain the FRB energy source. We then predict the host galaxy demographics for FRBs under different assumptions about magnetar formation channels, and demonstrate that this can be used as a powerful discriminant with $\lesssim 10$ precise localizations of FRBs (\S\ref{s:hosts}). Finally, we summarize our findings and present our conclusions (\S\ref{s:conclusions}).

\section{Magnetar Birth Rates}
\label{s:rates}

We begin by deriving volumetric rates, averaged over the redshift interval $0<z<0.5$, for the various transients that may be associated with magnetar formation. The choice of redshift interval is set by the observed dispersion measures (DMs) of FRBs, taking into account potential contributions from the host galaxy and/or FRB local environment (see \S\ref{s:frbs}). For SLSNe and GRBs, the remnant is assumed to be a \emph{millisecond magnetar}, as opposed to a slowly spinning classical magnetar that may be formed by a normal CCSN.

\subsection{Millisecond Magnetars}

\subsubsection{Superluminous Supernovae}

Energization by a nascent millisecond magnetar is now the prevailing model for Type I (hydrogen-poor) SLSNe \citep[see most recently][]{nic2016c, nic2017, nic2017b, kan2016b, ins2016b}. The rates of SLSNe at various redshifts between $0<z<3$ have been estimated by \citet{qui2013}, \citet{coo2012}, \citet{mcc2015}, and \citet{pss+.2017}. \citet{pss+.2017} normalized these rate estimates to the cosmic star-formation history derived by \citet{hb.2006, hb.2008}, who found SFR $\propto (1+z)^{3.28}$. The normalization is fixed to the estimated Type I SLSN rate at $z=0.17$, which is $32_{-26}^{+77}$\,Gpc$^{-3}$\,yr$^{-1}$, from \citet{qui2013}. This gives an SLSN rate $R_{\rm SLSN}(z)=19 (1+z)^{3.28}$\,Gpc$^{-3}$\,yr$^{-1}$. Averaging this function over the interval $0<z<0.5$ results in an estimated $\langle R_{\rm SLSN} \rangle \approx 40$\,Gpc$^{-3}$\,yr$^{-1}$. Due to the uncertainties in the individual rate measurements, the uncertainty on this estimate is at least a factor of 2.

SLSNe are found almost exclusively in low-luminosity (\apx0.01--0.5~$L_*$), metal-poor, star forming galaxies \citep{lun2014, lel2015, chen2016, per2016, ang2016,schu2016}. The host galaxy of \repeater\ is in fact typical of SLSN hosts \citep{mbm.2017}.

\subsubsection{Long Gamma-Ray Bursts}

LGRBs are also thought to be powered by central engines, but in this case the engine could be either a rapidly accreting black hole\footnote{For H-poor SLSNe, the necessary engine timescale exceeds the fallback time to the black hole \citep{dex2013}.} or a millisecond magnetar (e.g., \citealt{Thompson+04,mgt+2011,maz2014}). For the rate, we use the analysis of \citet{wp.2010} who infer a rate function of $R_{\rm LGRB}(z)=1.3(1+z)^{2.1}$\,Gpc$^{-3}$\,yr$^{-1}$ for $z<3$. Averaging over $0<z<0.5$ and using their beaming factor of $\approx 50$, we find $\langle R_{\rm LGRB}\rangle\approx 100$ Gpc$^{-3}$ yr$^{-1}$.  As in the case of the SLSNe, the uncertainty on this estimate is at least a factor of 2. Whether LGRBs form millisecond magnetars or black holes is an important consideration, as black holes are not expected to contribute to the FRB rate; we will consider both extreme cases (all LGRBs form magnetars or all form black holes) when determining the overall plausible range in birth rate in \S\ref{s:sum}.

The host demographics for LGRBs are similar to those for SLSNe, with exclusively star-forming galaxies and a significant preference for metal-poor dwarf galaxies ($\lesssim L_*$) (e.g., \citealt{chr2004,fru2006,schulze2015,perley2016}).

\subsubsection{Short Gamma-Ray Bursts}

Short gamma-ray bursts (SGRBs) are argued to result from the mergers of binary neutron stars (BNS), usually leading to the formation of black holes \citep{ber2014}.  However, some of these mergers may form stable millisecond magnetars, with the fraction depending on the masses of the neutron stars and the equation of state (e.g. \citealt{Metzger+08,Giacomazzo&Perna13}).  We stress that this scenario for a possible relation between SGRBs and FRBs is distinct from models which argue that the BNS mergers themselves are the sources of prompt cataclysmic FRBs \citep{tot2013}.

\citet{wp.2015} estimated the rates of SGRBs using data from BATSE, \emph{Swift}, and \emph{Fermi}. Their parameterization takes the form $R_{\rm SGRB}(z)=45 \exp[(z-0.9)/0.39)]$ Gpc$^{-3}$\,yr$^{-1}$, giving $\langle R_{\rm SGRB}\rangle\approx 270$\,Gpc$^{-3}$ yr$^{-1}$ at $0<z<0.5$; here we use a beaming factor of $\approx 30$ \citep{fong2015}. Therefore SGRBs could contribute significantly to the FRB progenitor birth rate if a large fraction make stable millisecond magnetars.  However, limits on the rates of extragalactic transients from radio time-domain searches \citep{met2015}, as well as late-time radio follow-up of SGRBs \citep{fong2016} suggest that at most a few percent of BNS mergers form stable millisecond magnetars.  This indicates that the SGRB formation channel is at most comparable and more likely sub-dominant to SLSNe and LGRBs.

In terms of host galaxy demographics, the SGRB rate depends on both stellar mass and star formation activity \citep{lei2010}, but there is no preference for low-metallicity galaxies similar to the hosts of \repeater, SLSNe, and LGRBs \citep{ber2009,fong.2013,ber2014}. SGRB hosts span a broad range of luminosities, \apx0.1--5\,$L_*$ \citep{ber2014}. \citet{fong.2013} found late- and early-type host galaxy fractions of \apx60--80\% and \apx20--40\%, respectively.

\subsubsection{Millisecond Magnetar Summary}
\label{s:sum}

We estimated rates for various exotic transients that may be associated with millisecond magnetar formation. The total millisecond magnetar birth rate depends on which of these transients are assumed to robustly produce millisecond magnetars. If only SLSNe leave behind millisecond magnetars, the birth rate is $\sim 40$\,Gpc$^{-3}$\,yr$^{-1}$. LGRBs can potentially increase this rate to $\sim 140$\,Gpc$^{-3}$\,yr$^{-1}$, while if $\sim 10$\% of BNS mergers also contribute to millisecond magnetar birth the rate may be $\sim 170$\,Gpc$^{-3}$\,yr$^{-1}$.  Given the relative uncertainties in the SLSN rate, the GRB beaming factors, and the BNS merger channel in general, the overall plausible range for the millisecond magnetar birth rate is $R_{\rm mag,ms} \approx$\,few\,$\times 10-100$\,Gpc$^{-3}$\,yr$^{-1}$. 

The relative contributions of the various possible channels will be imprinted in the host galaxy demographics.  We return to this point in \S\ref{s:hosts}.

\subsection{Classical Magnetars from Core-collapse Supernovae}
\label{s:ccsne}

Based on observations of Galactic magnetars, \citet{kou1998} and \citet{gill2007} estimate that \apx10\% of CCSNe form magnetars. It is unclear to what extent this population of magnetars differs from those in dwarf galaxies, which may be responsible for powering SLSNe and LGRBs. The most likely difference is that classical magnetars lack the rotational energy to significantly enhance the explosion energy or luminosity of the associated CCSNe, as they do in the case of SLSNe and LGRBs.  Observationally, this diversity seems to be connected to the pre-explosion environment: SLSNe and LGRBs show a strong preference for low metallicity, whereas normal CCSNe primarily follow the cosmic star formation rate distribution. Therefore it appears that magnetar birth may be different at low metallicity.

\citet{dah2004} derive an average CCSN rate at $0.1<z<0.5$ of $R_{\rm CCSN}\approx 2.5\times10^{5}$\,\pergpcperyr.  This implies a magnetar formation rate from CCSNe of $R_{\rm mag,CC}\approx 2.5\times 10^{4}$\,\pergpcperyr.  Therefore normal CCSNe rather than SLSNe or GRBs dominate the overall rate of magnetar production in the Universe by a factor of $>100$; this is also true in low metallicity galaxies.  One implication of this is that the vast majority of magnetars must be born with spin periods $\gg$1~ms to avoid an overproduction of SLSNe and LGRBs.

\section{Fast Radio Bursts: Rates and Characteristics}
\label{s:frbs}

For the observed FRB rate, we adopt the high-latitude event rate of \citet{vbl+.2016}, under the assumption that the low-latitude FRB rate is attenuated by Galactic propagation effects. This rate is $\mathcal{R}({>}S_{\nu,\text{lim}}) = 2870^{+4460}_{-1750}$ events per day across the sky at an observed flux density limit of $S_{\nu,\text{lim}}\ge 1$ Jy. The uncertainty bounds denote this number's 95\% confidence limit. This rate is somewhat higher than that inferred by \citet{lvl+.2016}, $\mathcal{R}=1320^{+840}_{-510}$~day$^{-1}$, due primarily to differences in estimated beam sizes \citep{vbl+.2016}. Because of this and other sources of uncertainty, the following calculation is therefore reliable only to within a factor of a few, similar to our estimates for the rate of magnetar-birthing transients.

\subsection{FRB search volume}
\label{s:frbsearchvol}

To convert the observed FRB arrival rate to a volumetric rate, we must first determine the volume within which we are sensitive to FRBs. To accomplish this, we consider the population of the 18 FRBs published at the time of writing. We adopt a simple model to estimate what portion of each known FRB's dispersion measure (DM) is intergalactic, then use the relation between redshift and DM in the intergalactic medium calculated by \citet{iok2003}.

To infer the intergalactic portion of each FRB's DM, we start with the observed DM values for all FRBs listed in FRBCat \citep{pet2016}. We subtract the contribution from the Milky Way disk derived from the NE2001 model \citep{cl.2002} and a fixed value of 30~pc~cm$^{-3}$ for the Milky Way halo contribution \citep{dgbb.2015}. For the FRB host galaxy contribution, we assume that the value inferred for \repeater\ \citep{tbc+.2017}, DM$_{\rm host}\approx 140$\,pc\,cm$^{-3}$, is typical of the population as a whole. X-ray observations of LGRBs at $z \lesssim 1$ reveal host hydrogen column densities $N_H \approx 10^{21}$--$5 \times 10^{22}$~cm$^{-2}$ \citep{swt+.2013}. Milky Way observations suggest that these values translate to a DM contribution of 3--150~pc~cm$^{-3}$ at an average ionization of 10\% \citep{hnk.2013}. \citet{lel2015} showed for SLSN hosts that the star-forming dwarf galaxies of interest tend to have high ionization parameters, which would result in higher DMs. Our adopted value is therefore plausible, but it is possible that our redshifts are biased somewhat low.

\begin{figure}
\includegraphics[width=\columnwidth]{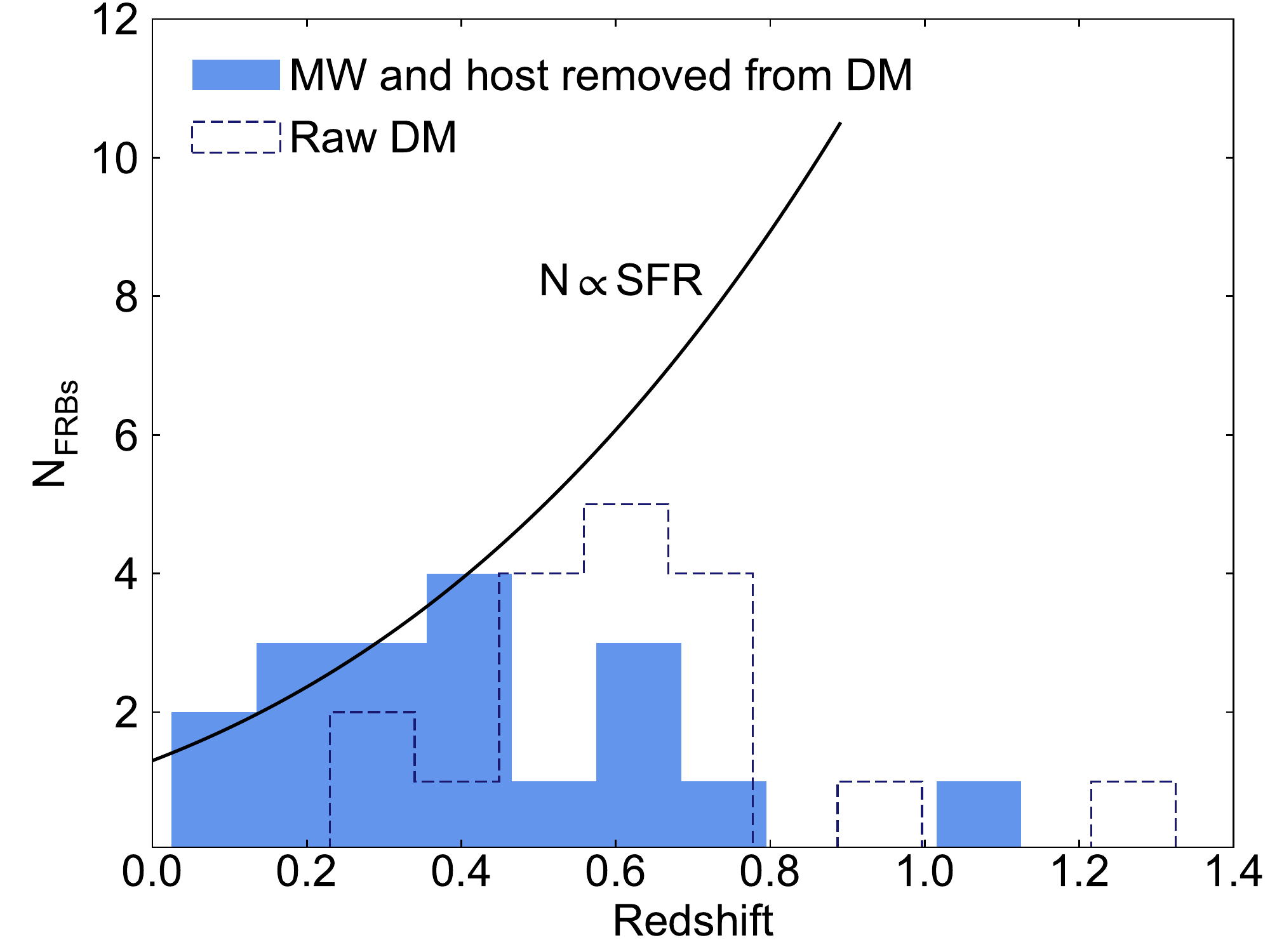}
\caption{The inferred redshift distribution of known FRBs, computed as described in \autoref{s:frbsearchvol}. The solid line shows the cosmic star-formation history (with arbitrary normalization). The distribution appears to be roughly complete only to $z\approx 0.5$. The dashed histogram shows the redshift distribution assuming that Milky Way and host contributions are zero (i.e., all DM is intergalactic). In this extreme model it remains the case that high-redshift (high-DM) FRBs are lacking.}
\label{f:redshifts}
\end{figure}

We then use the redshift-DM relation to find the redshift distribution of FRBs, which we show in \autoref{f:redshifts}. The distribution appears to decline sharply at $z\gtrsim 0.5$ and we therefore adopt $z<0.5$ as our fiducial search volume.

Bright FRBs should in principle be easily detectable at $z>0.5$, so the lack of events at higher estimated redshifts is surprising. This lack is especially striking if one expects the FRB production rate to scale with the cosmic star formation rate, shown as a black line in \autoref{f:redshifts}. Assessing whether this is a real physical effect associated with the source population, or an uncorrected observational bias, is beyond the scope of this paper.

\subsection{A Luminosity Function for FRBs from \repeater}

We compute the volumetric birth rate of repeating FRBs under the assumption that all FRBs are repeaters that share the same intrinsic luminosity density function as \repeater. We note that no repetitions have yet been detected from other FRBs. This is despite long term follow-up observations of some events with the Parkes Telescope \citep{petroff2015b,Ravi+15}. However, only the brightest burst from \repeater\ would likely have been detectable with Parkes \citep{ssh+.2016,sch2016}, so repetition of other sources cannot be excluded.

Observations of \repeater\ indicate intermittent periods of high activity (with many bursts) interspersed with periods in a low-activity state. In particular, \citet{clw+.2017} detected no bursts in 50 hours of VLA data in November 2015 and April-May 2016, before finding nine bursts in 33 hours in August-September 2016. The source therefore seems to spend about 30\%\ of the time in an active state. We denote this active duty cycle $\zeta$.

\begin{figure}
\includegraphics[width=\columnwidth]{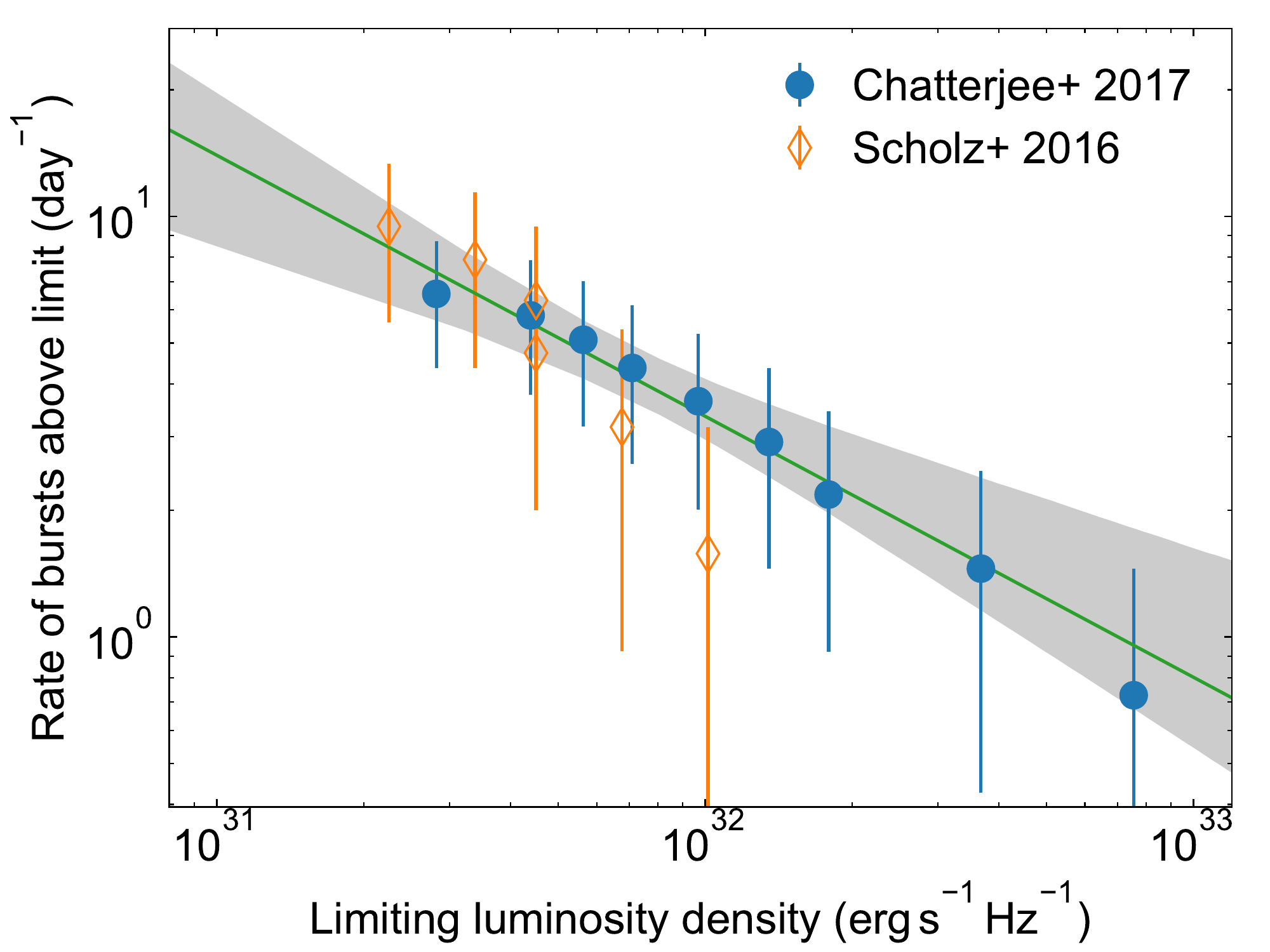}
\caption{Rate of observed bursts from \repeater\ (VLA, 3\,GHz) detected by \citet{clw+.2017}, as a function of minimum burst luminosity density. Error bars assume Poisson statistics. Our best fitting power law, with slope $\alpha = -0.62$, is marked (\autoref{e:powlaw}) along with the $1\sigma$ contours. Also shown for comparison are six bursts from \citet{sch2016} using the Green Bank Telescope at 2\,GHz.}
\label{f:powlaw}
\end{figure}

The nine VLA detections of \repeater\ published in \citet{clw+.2017} can be used to determine a fiducial cumulative distribution function for the isotropic (not beaming-corrected) luminosity density of individual FRB events, shown in \autoref{f:powlaw}. This function, $r({>}\Lnil)$, specifies the number of FRBs emitted per day, while the source is active, that are brighter than a threshold isotropic luminosity density \Lnil. The errors in each bin of our empirical cumulative distribution are assumed to be dominated by Poisson noise. We normalize the distribution to a total observing time of 33 hours; the actual observed rate is thus $\zeta r({>}\Lnil)$, where $\zeta \approx 0.3$ based on \citet{clw+.2017}.  The luminosity function is well modeled by a power law of the form:
\begin{equation}
r({>}\Lnil) = k \left(\frac{\Lnil}{10^{32}\text{ erg s}^{-1}\text{ Hz}^{-1}}\right)^\alpha,
\label{e:powlaw}
\end{equation}
with $\alpha = -0.62^{+0.18}_{-0.22}$, and $k = 3.5^{+0.6}_{-0.6}$ day$^{-1}$. 

With this luminosity function and duty cycle, and for the redshifts and search times of the FRBs followed up by \citet{petroff2015b} and \citet{Ravi+15}, no such FRB source has an expectation of $>0.5$ repetitions above the limiting sensitivity of Parkes \citep[$\approx 0.15$\,Jy;][]{sch2016}, confirming that the currently known FRBs are consistent with also being repeaters. \citet{lu2016} estimated a similar distribution function for \repeater\ (but for burst energy rather than luminosity) and also found that if this function is universal, repetition from other FRBs is not excluded by the existing data.

We denote the beaming factor $\eta$, such that the true luminosity density of each burst is $L_\nu = \eta L_{\nu,\text{iso}}$. The value of $\eta$ for FRBs is not constrained observationally. We adopt a fiducial value of $\eta = 0.1$, appropriate for pulsars \citep{tm.1998}.

\subsection{The FRB source density}

We denote the volumetric birth rate of repeating FRBs as $R_{\rm FRB}(z)$, measured in units of \pergpcperyr, and the lifetime of each repeater as $\tau$~yr. The comoving volume density of repeaters is therefore $R_{\rm FRB}\tau$. We assume that $R_{\rm FRB}(z)$ follows cosmic star-formation history \citep{hb.2006,hb.2008}, defining $R_{\rm FRB}(z) = R_0 (1+z)^{3.28}$. While the star formation rate (for a given mass) is observed to be greater than average in galaxies with the low metallicities appropriate to \repeater\ and SLSNe, any differential evolution with redshift compared to normal star-formation appears to be negligible within the small redshift range used in this study \citep[e.g.][]{mannucci2010}. The observer frame birth rate and lifetime are both affected by time dilation but in an opposite sense, so this effect cancels in the product $R_{\rm FRB}\tau$.  We will explore constraints on $\tau$ in the next section.

In this framework the observed rate of FRBs on the sky is
\begin{equation}
\mathcal{R} = \int_0^{0.5} \dd z \frac{\dd V}{\dd z} \eta \zeta r({>}4 \pi d_L^2(z) S_{\nu,\text{lim}}) R_0 (1 + z)^{2.28} \tau.
\label{e:intrate}
\end{equation}
Here we take into account FRBs beamed away from the observer ($\eta$), the active duty cycle of each FRB source ($\zeta$), the effects of surveying to fixed flux density sensitivity ($\Lnil = 4\pi d_L^2(z) S_{\nu,\text{lim}}$), and the effects of time dilation upon the observed FRB rate (a factor of $(1 + z)^{-1}$ folded into the cosmic star formation rate history). We neglect any effects of redshift upon the observed FRB pulses (i.e., $k$-corrections).

Using our power-law model (\autoref{e:powlaw}) and the observed value of $\mathcal{R}$, we evaluate the integral numerically and find $R_0 \tau \approx 130\, (\eta \zeta)^{-1}$\,Gpc$^{-3}$. The volume density of repeating FRB sources averaged over $0 < z < 0.5$ is $R_{\rm FRB}\tau \approx 280\, (\eta \zeta)^{-1}$\,Gpc$^{-3}$. We derive the same density (within 10\%) if we assume a uniform (rather than star-formation--weighted) distribution, demonstrating that our estimate is not sensitive to the choice of star-formation history (e.g.~as a function of metallicity). The uniform distribution would also be more appropriate for BNS mergers. Taking $\eta = 0.1$ and $\zeta = 0.3$, the fiducial volume density of repeaters is therefore $R_{\rm FRB} \tau \sim 10^4$\,Gpc$^{-3}$.

\citet{lu2016} estimated the volume density of repeaters as $\sim 10^2-10^4$\,Gpc$^{-3}$, where the wide range comes from the uncertainty in the all-sky rate at that time, but did not account for the beaming or duty cycle (i.e.~$\eta=1$ and $\zeta=1$). Given these differences, their estimate agrees well with ours.

\subsection{Comparison of the Magnetar Birth Rate and FRB Rate}

If most millisecond magnetars are FRB sources and this is the dominant FRB production channel, we have $R_\text{FRB} \approx R_\text{mag,ms}$. In this scenario, the lifetime of a typical repeater must be \apx30--300~yr.  The lower limit is reassuring from the perspective that supernova ejecta generically require \apx10~yr to become optically thin at GHz frequencies \citep{mbm.2017}, thus allowing FRB emission to escape to the observer. This lifetime is also compatible with \repeater, which has so far been active for \apx4~yr.

If slowly-rotating classical magnetars from CCSNe produce FRBs, these would necessarily dominate the rate (\S\ref{s:ccsne}). Setting $R_{\rm FRB} = R_{\rm mag,CC}$, we find $\tau \lesssim 0.5$\,yr. Even if we take only those CCSNe in galaxies of comparable luminosity to the host of \repeater\ \citep{sve2010}, we find $\tau < 5$\,yr. This timescale is short compared to the time for the ejecta to become transparent and is only marginally compatible with \repeater. The essential point is that if most FRBs are repeaters, at most a small minority of CCSNe ($\ll 1$\%) can make FRB sources in order for the rates and timescales to be consistent with observations, whether or not one assumes that this minority is indeed the SLSNe and LGRBs.

\subsection{Energetics}

In the previous section we derived a power-law estimate for the rate of repeater bursts as a function of their isotropic luminosity density. We can use this power law to determine the rate at which the repeater is losing energy, which can be compared to the various mechanisms that may power FRB emission.

The energy emitted per individual FRB event is
\begin{equation}
E = \eta L_{\nu,\text{iso}} \Delta t \Delta \nu,
\end{equation}
where $\Delta t$ and $\Delta \nu$ are the characteristic durations and bandwidths of each event. We assume $\Delta t = 1$~ms and $\Delta \nu = 1$~GHz. By the definition of the FRB rate function $r({>}\Lnil)$, when an FRB source is active it produces bursts within an infinitesimal isotropic luminosity density interval $[L_{\nu,\text{iso}}, L_{\nu,\text{iso}} + \Delta L_{\nu,\text{iso}}]$ at a rate of
\begin{equation}
|r'(L_{\nu,\text{iso}})| \Delta L_{\nu,\text{iso}},
\end{equation} 
where $r'$ is the derivative of $r$ with respect to $L_{\nu,\text{iso}}$. The rate of energy loss due to FRBs is therefore
\begin{equation}
\dot E = \int_{L_{\nu,\text{iso,min}}}^{L_{\nu,\text{iso,max}}} \dd L_{\nu,\text{iso}} \, E \zeta |r'(L_{\nu,\text{iso}})|,
\end{equation}
where the duty cycle of FRB production has now been factored in. Evaluating the integral over a finite range in $L_\nu$ we find
\begin{align}
\dot E &= 
  \left|\frac{\alpha}{1 + \alpha}\right|
  k \eta \zeta \Delta \nu \Delta t
  \frac{L_{\nu,\text{iso,max}}^{\alpha + 1} - L_{\nu,  
    \text{iso,min}}^{\alpha + 1}}{L_{\nu,0}^\alpha} \nonumber \\
& \gtrsim 4 \times 10^{32}
  \left(\frac{\Delta\nu}{1\text{ GHz}}\right)
  \left(\frac{\Delta t}{1\text{ ms}}\right)
  \left(\frac{\eta}{0.1}\right)
  \left(\frac{\zeta}{0.3}\right)
  \text{ erg s}^{-1},
\label{e:edot}
\end{align}
where $L_{\nu,0} = 10^{32}$~erg~s$^{-1}$~Hz$^{-1}$ (see \autoref{e:powlaw}). In the adopted power-law model for $r$ where $\alpha \approx -0.7$, the value of the integral in \autoref{e:edot} is only weakly dependent on $L_{\nu,\text{iso,min}}$, and energy loss is dominated by the rare, luminous bursts close to $L_{\nu,\text{iso,min}}$\footnote{In constrast, for a steeper power-law ($\alpha<-1$), constraints on the faint end of the luminosity function would be more important.}. Therefore integrating over the observed range in $L_{\nu,\text{iso}}$ up to the brightest observed burst (\autoref{f:powlaw}) yields a constraining lower limit on the total dissipation.
We find that the total energy emitted in FRBs over 100\,yr is $E_{\rm tot} \gtrsim 10^{42}$\,erg, assuming fiducial parameters.

\citet{mbm.2017} suggest that the FRB energy reservoir is either the rotational or magnetic energy of the magnetar. They estimate an internal magnetic energy $E_{\rm B}\sim10^{49}$\,erg. The total emitted energy in magnetically-powered FRBs should be roughly this value multiplied by some efficiency factor for making coherent radio emission. For a lifetime of 30--300~yr, the required efficiency factor is $E_{\rm tot}/E_{\rm mag} \sim 10^{-7.5}\text{--}10^{-6.5}$.  This is on the high end of theoretical predictions for the efficiency of GHz radio emission from synchrotron maser instability for a magnetic pulse interacting with the magnetar nebula \citep{lyu2014}.  This may support alternative models such as ultra-relativistic internal shocks between mass ejection between subsequent flares \citep{bel2017}.

In the case of rotational powering, the available energy is 
\begin{equation}
E_{\rm rot}(t) \approx 2 \times 10^{52} P_{\rm ms}^{-2} (1+ t/\tau_{\rm sd})^{-1}\,{\rm erg},
\end{equation}
where $P_{\rm ms}$ is the initial spin period in milliseconds and $\tau_{\rm sd}$ is the spin-down timescale \citep{ost1971,kas2010}. For the typical values inferred for SLSNe ($P_{\rm ms} \gtrsim 1$, $\tau_{\rm sd} \gtrsim$\,days; \citealt{nic2017b}), the rotational energy remaining at $t \gtrsim 10$\,yr is $E_{\rm rot} \sim 10^{47}-10^{49}$\,erg. In the case of LGRBs, the spin-down time is $\ll 1$\,day, in which case $E_{\rm rot} \sim 10^{46}\text{--}10^{47}$\,erg. If FRBs are powered by rotational energy, the required efficiency is uncomfortably high \citep{Lyutikov17,mbm.2017}, especially in the case of LGRBs.  Thus magnetic powering may be the more likely scenario.  Alternatively, this may suggest SLSNe are a more suitable channel for producing FRBs than are LGRBs.

\subsection{Quiescent sources}

Given our constraints on the volumetric rate of FRB sources in the magnetar model, we can also ask what this implies about detecting their quiescent radio sources with future continuum wide-field radio surveys, assuming they are all similar to those of FRB 121102.

The radio luminosity of the quiescent source associated with FRB 121102 was $L_{\nu} \approx 2\times 10^{29}$ erg s$^{-1}$ Hz$^{-1}$ at 1.7$-$5 GHz.  The number of all-sky sources which would be detectable to a given flux depth $F_{\nu, \rm lim}$, assuming Euclidean geometry, is given by
\begin{align}
&N_{\rm all-sky} = \frac{4\pi}{3}\left(\frac{L_{\nu}}{4\pi F_{\nu, \rm lim}}\right)^{3/2}{R}_{\rm FRB}\tau  \nonumber \\
&\approx 10^3\left(\frac{L_{\nu}}{10^{29}\,{\rm erg}\,{\rm s}^{-1} {\rm Hz}^{-1}}\right)^{3/2} \left(\frac{F_{\nu, \rm lim}}{\rm 1\,mJy}\right)^{-3/2} \left(\frac{{R}_{\rm FRB}\tau }{10^{4}\,{\rm Gpc^{-3}}}\right)
\end{align}

Comparing to Figure 3 of \citet{Metzger+15b}, we see that the all-sky rate of quiescent sources exceeds all explosive transient types in this frequency range (e.g.~orphan LGRB afterglows, jetted tidal disruption flares), with the possible (if speculative) exception of stable millisecond magnetars from BNS mergers --- which, as we have discussed, could indeed be one source of FRBs.  

Upcoming wide-field surveys such as the Very Large Array Sky Survey (e.g., \citealt{Murphy+15}) and the VAST survey on ASKAP (\citealt{Murphy+13}) could in principle detect thousands of these quiescent radio sources.  However, if they are powered by the same mechanism giving rise to the actual FRB flares, their flux will evolve slowly, on timescales of decades to centuries, comparable to the age of the source or the FRB active duration.  Still, over the duration of a 3-4 year survey it is conceivable that some sources would show a detectable fading.  This behavior might be challenging to distinguish from e.g.~variable AGN, though an association with dwarf galaxies would help in this regard. One could also look for these quiescent sources directly in galaxies that have previously hosted known SLSNe or GRBs, or indeed BNS mergers if they are detected via gravitational waves.

\subsection{Cataclysmic FRBs?}
\label{s:cataclysmic}

In this section we investigate the volumetric FRB rate under the alternative assumption that the FRB rate is dominated by cataclysmic events, i.e.~where each source emits only a single burst in its lifetime, rather than \repeater-like repeaters. In this case, we lack a fiducial luminosity function, but we can set a lower bound by assuming that all FRBs within $z<0.5$ are brighter then 1\,Jy, leading to:
\begin{equation}
\mathcal{R} = \int_0^{0.5} \dd z \frac{\dd V}{\dd z} \eta R_{0{\rm , cat}} (1+z)^{2.28},
\end{equation}
where $R_{0{\rm , cat}}$ is the volumetric rate of cataclysmic FRBs at $z=0$ and we have again assumed that the FRB rate scales with the star formation rate density and accounted for beaming and time dilation. $\eta$ is the beaming factor as before\footnote{In the case of cataclysmic FRBs, there is no need to consider a duty cycle, as their `lifetime' is simply the duration of one burst.}.

We find that the required volume-averaged cataclysmic FRB rate is $R_{\rm cat} \approx 3.6\times 10^4 \eta^{-1}$\,\pergpcperyr. In this case FRBs cannot be associated with the birth of millisecond magnetars.  They would need to come from a much more common process than SLSNe and GRBs. The rate we derive is about $(10/\eta)\%$ of the CCSN rate \citep{dah2004}, implying that FRBs could be associated with the formation of classical magnetars if the spin period is irrelevant to FRB production\footnote{The rate of magnetic field evolution inside a neutron star is a strongly increasing function of its interior magnetic field strength and mass (e.g.~\citealt{Beloborodov&Li16}). If the millisecond magnetars associated with SLSNe or LGRBs have significantly larger masses (e.g. due to their origin from very massive stars) or higher interior fields (e.g. due to their rapid birth rotation) than classical magnetars, this could result in them possessing an enhanced rate of FRB activity in magnetically-powered scenarios (\citealt{bel2017}).}, FRBs are not absorbed by CCSN ejecta, and FRBs are only mildly beamed. However, this scenario has distinct implications for the host galaxy demographics that are in tension with the localization of \repeater\ to a dwarf galaxy, as we explore in detail in the next section. For our adopted value of $\eta$, the cataclysmic FRB rate would have to be comparable to the CCSN rate.

\section{Host Galaxy Demographics}
\label{s:hosts}

The fact that \repeater\ resides in a dwarf galaxy ($L\approx 0.02 L_*$) would be very unlikely given any scenario in which the FRB rate follows stellar mass or the cosmic star formation rate. In this section, we address this quantitatively, and investigate what we can learn from future localizations.

It is well established that hydrogen-poor SLSNe show a strong preference for dwarf galaxies \citep{nei2011,lun2014,lel2015,ang2016,schu2016}. These show many properties in common with LGRB hosts \citep{lun2014}, including an apparent metallicity threshold \citep{chen2016,per2016}, though this threshold appears to be lower for SLSNe than for LGRBs \citep{schu2016}.  Almost all SLSNe and most LGRB hosts have $B$-band absolute magnitudes in the range $-15\gtrsim M_B\gtrsim -20$ (i.e., $<0.5 L_*$), and these galaxies have low metallicities \citep{chen2013,lun2014}. 

The host galaxy of \repeater\ is remarkably consistent with these properties \citep{mbm.2017}.  \citet{tbc+.2017} determine an extinction-corrected absolute magnitude $M_r=-17$, a metallicity upper limit $12+\log ([{\rm O/H}])\lesssim 8.4$, and emission line equivalent widths consistent with extreme emission line galaxies, such as SLSN hosts \citep{lel2015,schu2016}.

Therefore if FRBs do indeed result from the millisecond magnetar remnants of SLSNe and possibly LGRBs, we would expect most FRBs to be found in galaxies similar to the host of \repeater.  Using the inferred volumetric rate of SLSNe and rate density of repeating FRBs, we can determine the fraction of dwarf galaxy hosts by calculating the number density of dwarf galaxies. This is achieved by integrating the observed galaxy luminosity function over the magnitude range relevant to SLSN/LGRB/FRB hosts.

\begin{figure}
\includegraphics[width=\columnwidth]{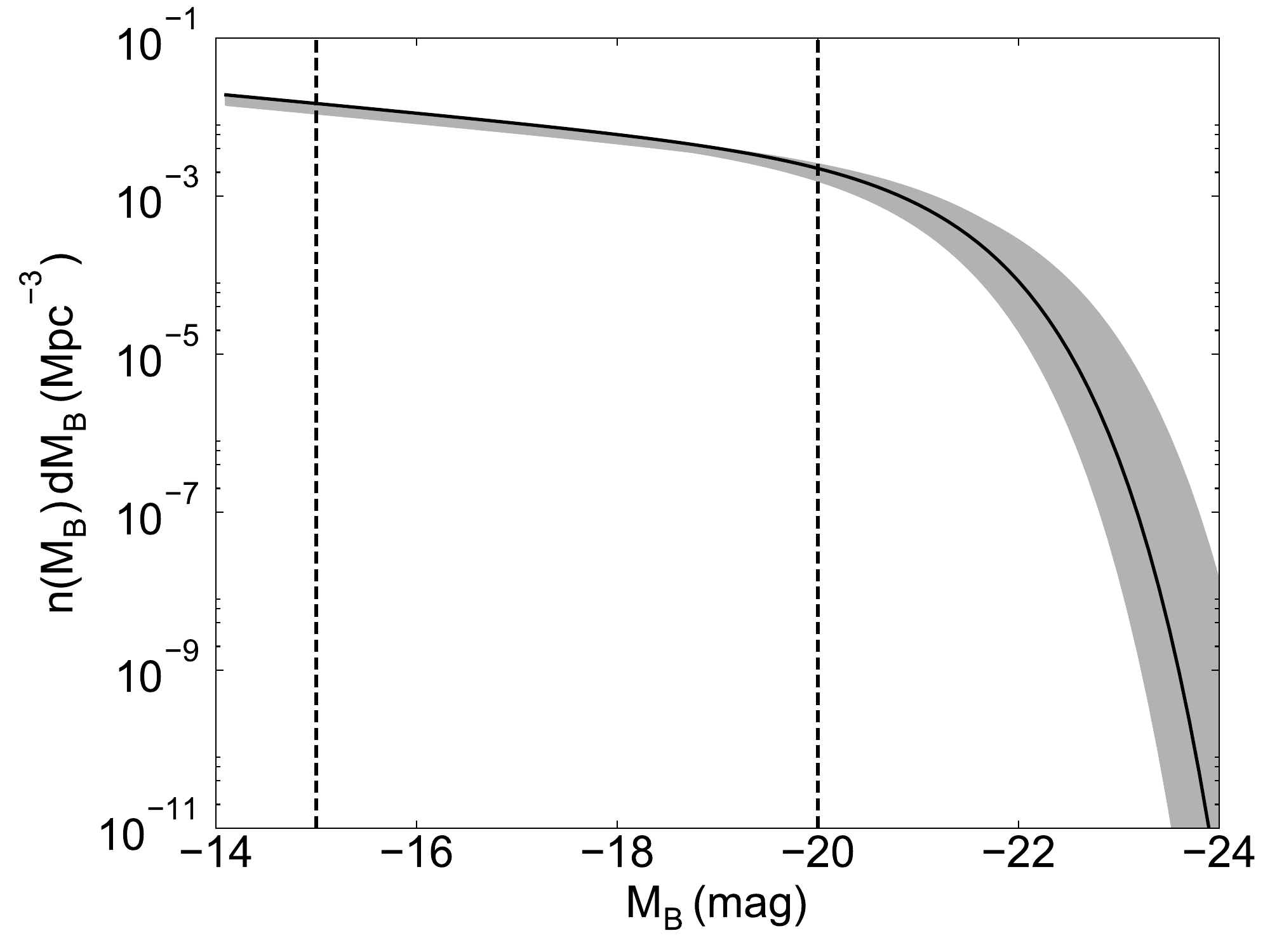}
\caption{The galaxy luminosity function from \citet{fab2007} for blue galaxies. The black line is our fiducial model ($z=0.5$), while the grey area shows the evolution of luminosity function at $0.3<z<1.1$. The evolution is negligible over the magnitude range of SLSN and LGRB host galaxies, marked by dashed vertical lines.}
\label{f:schechter}
\end{figure}

\citet{fab2007} measure the galaxy luminosity function separately for red and blue galaxies, in several redshift bins at $0.3<z<1.3$. The luminosity function for blue, star-forming galaxies is appropriate in this case. Over the magnitude and redshift range of interest, the number density of dwarf galaxies shows negligible evolution with redshift (\autoref{f:schechter}), we therefore take as fiducial values their luminosity function at $z=0.5$. Integrating over $-15>M_B>-20$, we find a dwarf galaxy number density of $3.0\times 10^7$\,Gpc$^{-3}$.

Taking our estimated SLSN rate of 40\,\pergpcperyr, this corresponds to a per-galaxy rate of $\sim 1$\,Myr$^{-1}$ averaged to $z\approx 0.5$.  Interestingly, assuming a progenitor lifetime of $\sim {\rm few}$\,Myr, appropriate for very massive stars, this implies the existence of $\sim 1$ SLSN progenitor in a typical dwarf galaxy like the LMC at any given time. If all FRBs are repeaters located in similar galaxies, our number density, $R_{\rm FRB}\tau \sim 10^4$\,Gpc$^{-3}$, implies that only a few in $10^5$ dwarf galaxies currently hosts a repeating FRB.

\begin{figure}
\includegraphics[width=\columnwidth]{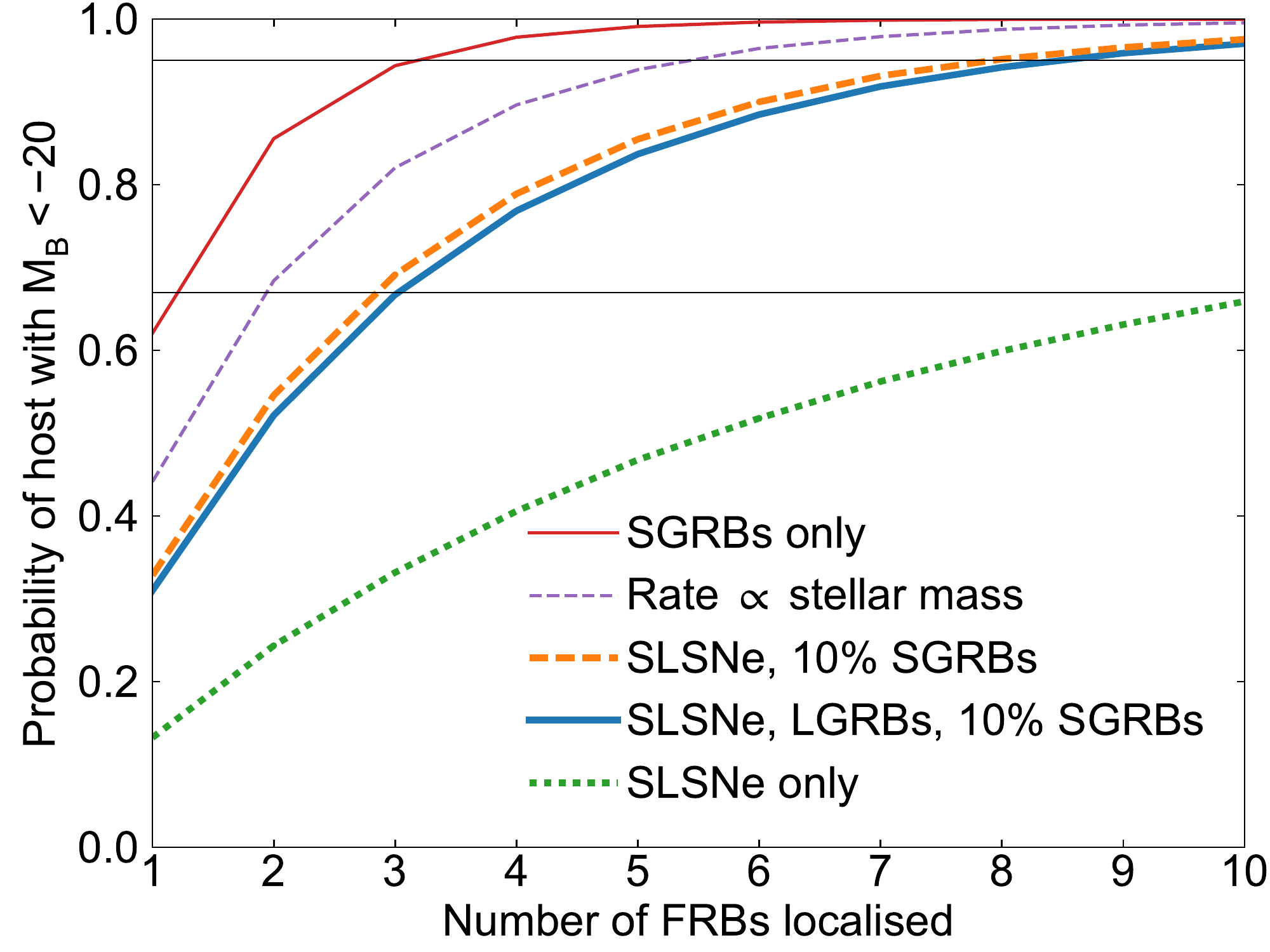}
\caption{
The probability of localising at least one FRB in a massive (as opposed to dwarf) galaxy as a function of the number of localised FRBs. The various curves represent different assumptions about which transients produce FRB sources, while the horizontal lines indicate 67\%\ and 95\%\ probability.
}
\label{f:host}
\end{figure}

While it is appealing to draw this link between FRBs and other rare transients in similar environments, it is important to realise that this argument currently rests on a single FRB localisation. Therefore a more useful exercise is to consider the number of FRBs that would need to be localised (and therefore associated with a host galaxy) to exclude a population in massive galaxies ($M_B<-20$, or $L>0.5L_*$) at a certain confidence level. We perform a series of Monte Carlo experiments for various FRB progenitor channels described below. In each case we simulate 10 FRB localisations, with the host galaxies randomly drawn from the expected fractions in dwarf and massive galaxies for each progenitor channel. We take the dwarf galaxy fractions (described for each transient population below) from the literature, and conservatively assume a Gaussian error of 10\% in each case. We repeat this process $10^5$ times to determine the cumulative probability of finding a massive FRB host galaxy as progressively more FRBs are localised.

If a large fraction of BNS mergers (observed as SGRBs) leave behind millisecond magnetars, these could dominate the FRB rate. \citet{ber2014} find that 38\%\ of SGRBs occur in galaxies with $M_B>-20$. Our simulations show that a mere 3 localisations in dwarf galaxies are needed to exclude at 95\% confidence a channel where only SGRBs form FRB sources.  Moreover, the SGRB channel predicts that a substantial fraction ($\sim 30$\%) of FRBs should occur in elliptical galaxies. This is in contrast to any supernova-related channel.

More generally, we can try assuming that the FRB rate traces the stellar mass density. This could be applicable to any formation channel that does not require a young stellar population (one example being the BNS mergers above). In this case we simply integrate the luminosity function from \citet{fab2007} below and above $M_B=-20$, finding that 44\% of stellar mass---and hence the same fraction of FRBs---should reside in galaxies brighter than the cutoff. These relative fractions are also consistent with expectations for CCSNe (though technically CCSNe trace star-formation rate rather than stellar mass).  With this ratio, we require $\sim 5$ localisations in exclusively dwarf galaxies to exclude such a channel.

We next consider the hypothesis that FRBs are formed by SLSNe and both types of GRBs. We take our fiducial rates from \S\ref{s:rates}, and use the following host demographics: 86\%\ of SLSNe and 70\%\ of LGRBs occur in galaxies with $M_B>-20$ \citep{lun2014}\footnote{At $z<0.5$, these fractions appear to be closer to 100\%, but we use the overall values to be conservative.}, while only 38\%\ of SGRBs do (as noted above). We find that $\approx 8$ localisations are needed to rule out (at 95\%\ confidence) a SGRB channel for producing FRBs that is comparable in rate to the SLSN and LGRB channels (as opposed to the SGRB-only channel investigated above, which required only 3 localisations), regardless of how much LGRBs contribute relative to SLSNe. 

Finally, in the SLSN-only scenario, virtually all FRBs are expected to be in galaxies with $M_B>-20$. Localising $\lesssim 10$ FRBs in exclusively dwarf galaxies would therefore allow us to robustly rule out any model with a significant contribution from SGRBs or any progenitors that trace stellar mass, and strongly favor SLSNe as the primary channel for FRB production.

\begin{deluxetable}{lr}
\tablecolumns{2}
\tablewidth{0em}
\tablecaption{Volumetric event rates, $0 < z < 0.5$\label{t:rates}}
\tablehead{
 \colhead{Event type} & \colhead{Rate} \\
   & \colhead{(\pergpcperyr)}
}
\startdata
Superluminous supernovae & 40 \\
Long gamma-ray bursts & 100 \\
Short gamma-ray bursts & 270 \\
Overall millisecond magnetar formation & few $\times$ 10--100 \\
Core-collapse supernovae & $2.5 \times 10^5$ \\
CCSNe forming classical magnetars & $2.5 \times 10^4$ \\
FRB repeater birth & $10^4 / \tau$ \\
Cataclysmic FRBs & $3.6 \times 10^5$ \\
\enddata
\tablecomments{Rates are averages over $0 < z < 0.5$ and adopt fiducial parameters of $\eta = 0.1$ and $\zeta = 0.3$. For FRB repeaters, $\tau$ is the mean repeater lifetime. All rates are uncertain to factors of at least a few.}
\end{deluxetable}

\section{Summary and Conclusions}
\label{s:conclusions}

We explored the possibility that repeating FRB sources are the millisecond magnetar remnants of superluminous supernovae and GRBs, by comparing volumetric rates and host galaxy demographics. We collect the key volumetric event rates in \autoref{t:rates}. Given the rarity of the sources involved (CCSNe not withstanding), the various estimates are accurate only to within a factor of a few. However, this is sufficient to discriminate between several possible FRB formation channels. Our main findings are:
\begin{itemize}

\item The number density of repeating FRB sources at $0<z<0.5$, which is the product of their birth rate and active lifetime, is $R_{\rm FRB} \tau\approx 10^4$\,Gpc$^{-3}$, accounting for a beaming factor of about 0.1 and an active duty cycle of about 0.3.

\item There are hints that high-redshift FRBs are lacking, especially if the FRB production rate tracks the cosmic star formation rate. In our adopted model (\autoref{e:intrate}), twice as many FRBs should be observed originating from $0.5 < z < 1$ as $0 < z < 0.5$, but this is not the case (\autoref{f:redshifts}). This result, however, is sensitive to assumptions about the FRB progenitor population, the relationship between DM and redshift, and the observational biases in current FRB surveys, which are not fully understood. Future work should investigate this matter more rigorously.

\item If FRBs are formed by CCSNe, at most $\approx 0.1$\%\ of CCSNe actually result in FRBs. The longer the FRB lifetime inferred from continued monitoring of \repeater, the smaller this fraction becomes. This fraction is comparable to the relative rate of SLSNe/LGRBs compared to CCSNe.

\item If SLSNe and LGRBs are the progenitors of millisecond magnetars that emit FRBs, observed FRB detection rates imply a repeater lifetime of \apx30--300~yr. This is consistent with \repeater, as well as the time it takes for the supernova ejecta to become transparent at GHz frequencies \citep{mbm.2017}. 

\item The integrated energy released by repeaters with such lifetimes is consistent with the high end of the range of expectations for magnetar-powered models.  Given theoretical estimates for the radio emission efficiency, magnetically-powered models may be preferable to rotationally-powered ones, particularly if LGRBs contribute to FRB formation.

\item A cataclysmic channel, in which most FRBs are single pulses rather than repeaters, requires a very high rate, $>10$\% of the CCSN rate. However, as we already noted, CCSN ejecta are opaque to FRBs at the time of explosion.  Additionally, the potential preference for dwarf galaxies may already disfavor a model in which the FRB progenitors are so common.

\item As pointed out by \citet{tbc+.2017}, the host galaxy of \repeater\ is remarkably similar to the hosts of SLSNe and LGRBs. Localizing 3--5 more FRBs in exclusively dwarf galaxies should be sufficient to exclude a dominant channel with a significant population in massive galaxies, such as SGRBs or any model that traces stellar mass. Finding \apx7 FRBs exclusively in dwarf galaxies would rule out a significant contribution from SGRBs.

\item Directly finding the quiescent sources associated with FRBs (rather than locating them by following up a specific burst) will be challenging, as we expect only one FRB source per $10^5$ galaxies, though in principle there may be thousands of such sources on the sky. Most dwarf galaxies should harbor \apx1 SLSN progenitor at any given time. A more efficient search strategy may be to target galaxies that have hosted SLSNe, GRBs, or gravitational wave sources (in the case of BNS mergers) in the past tens of years. This is currently feasible for many GRBs and for the oldest known SLSNe.
\end{itemize}

If \repeater\ is a typical FRB source, then the overall properties of FRBs (in terms of rates, lifetimes, energetics and host galaxies) are consistent with expectations for millisecond magnetars from SLSNe and/or LGRBs. Additional precise localizations of FRBs will stringently test our proposed scenario, with \apx10 localizations needed to provide valuable insight into the nature of FRBs from their host galaxy demographics. Another important empirical task is to test if the low number of FRBs at inferred redshifts $z > 0.5$ reflects a sharp cutoff, and if so, whether this effect is astrophysical or observational.

\acknowledgments \textit{Acknowledgments.}
The Berger Time-Domain Group is supported in part by NSF grant AST-1411763 and NASA ADA grant NNX15AE50G.
BDM gratefully acknowledges support from the National Science Foundation (AST-1410950, AST-1615084), and NASA through the Astrophysics Theory Program (NNX16AB30G). 
This research has made use of NASA's Astrophysics Data System.

\bibliographystyle{yahapj}
\bibliography{references}

\end{document}